\documentclass[runningheads]{llncs}

\usepackage{eccv}

\usepackage{eccvabbrv}

\usepackage{graphicx}
\usepackage{booktabs}
\usepackage{amsmath,scalerel}
\usepackage{amssymb}
\usepackage{amsfonts}       
\usepackage{nicefrac}       
\usepackage{microtype}      
\usepackage{mathtools}
\usepackage{caption}
\usepackage{color, colortbl}
\usepackage{soul}
\usepackage{graphics,subcaption}
\usepackage{adjustbox}
\usepackage{multirow}
\usepackage{diagbox}
\usepackage{tikz,siunitx}
\usepackage[ruled,vlined]{algorithm2e}
\usepackage{wrapfig}
\usepackage[normalem]{ulem}
\useunder{\uline}{\ul}{}

\usepackage[accsupp]{axessibility}  

\usepackage{hyperref}

\usepackage{orcidlink}

\begin{document}

\title{Architecture-Agnostic Untrained Network Priors for Image Reconstruction with Frequency Regularization} 

\titlerunning{Architecture-Agnostic Untrained Network Priors }

\author{Yilin Liu\inst{1}\orcidlink{0000-0002-2540-1295} \and
Yunkui Pang\inst{1}\orcidlink{0000-0003-2798-337X} \and
Jiang Li\inst{1}\orcidlink{0009-0003-6259-4932} \and
Yong Chen\inst{2}\orcidlink{0000-0001-6183-2693} \and
Pew-Thian Yap\inst{1}\orcidlink{0000-0003-1489-2102}}

\authorrunning{Liu et al.}

\institute{Computer Science, University of North Carolina at Chapel Hill \and
Radiology, Case Western Reserve University \\
\email{\{yilinliu,yunkuipa,jianglz\}@cs.unc.edu, yxc235@case.edu, ptyap@med.unc.edu}
}

\maketitle

\begin{abstract}
  Untrained networks inspired by deep image priors have shown promising capabilities in recovering high-quality images from noisy or partial measurements \textit{without requiring training sets}. Their success is widely attributed to implicit regularization due to the spectral bias of suitable network architectures. However, the application of such network-based priors often entails superfluous architectural decisions, risks of overfitting, and lengthy optimization processes, all of which hinder their practicality. To address these challenges, we propose efficient architecture-agnostic techniques to directly modulate the spectral bias of network priors: 1) bandwidth-constrained input, 2) bandwidth-controllable upsamplers, and 3) Lipschitz-regularized convolutional layers. 
  We show that, with \textit{just a few lines of code}, we can reduce overfitting in underperforming architectures and close performance gaps with high-performing counterparts, minimizing the need for extensive architecture tuning. This makes it possible to employ a more \textit{compact} model to achieve performance similar or superior to larger models while reducing runtime. Demonstrated on inpainting-like MRI reconstruction task, our results signify for the first time that architectural biases, overfitting, and runtime issues of untrained network priors can be simultaneously addressed without architectural modifications. Our code is publicly available \footnote{https://github.com/YilinLiu97/Untrained-Recon.git}. 
\end{abstract}

\section{Introduction}
Magnetic resonance imaging (MRI) is a mainstream imaging tool for medical diagnosis. Reconstructing MR images from raw measurements involves data transformation from a Fourier spectrum in \(k\)-space to image space \cite{hansen2015image,liu2021real}. Since acquiring full \(k\)-space measurements is time-consuming, under-sampled \(k\)-space data are often collected to reduce scan times. This makes accelerated MRI reconstruction an ill-posed inverse problem that conventionally requires handcrafted priors \cite{lustig2007sparse,lingala2011accelerated} to mitigate the resulting aliasing artifacts. While supervised learning methods based on convolutional neural networks (CNNs) have enhanced reconstruction quality with fewer measurements, their training relies on paired under-sampled and fully-sampled measurements, which are expensive to acquire and can affect robustness and generalization across different acquisition protocols or anatomical variations  \cite{knoll2019assessment,knoll2020advancing}.

\begin{figure}[t!]
    \centering
    \includegraphics[width=\linewidth]{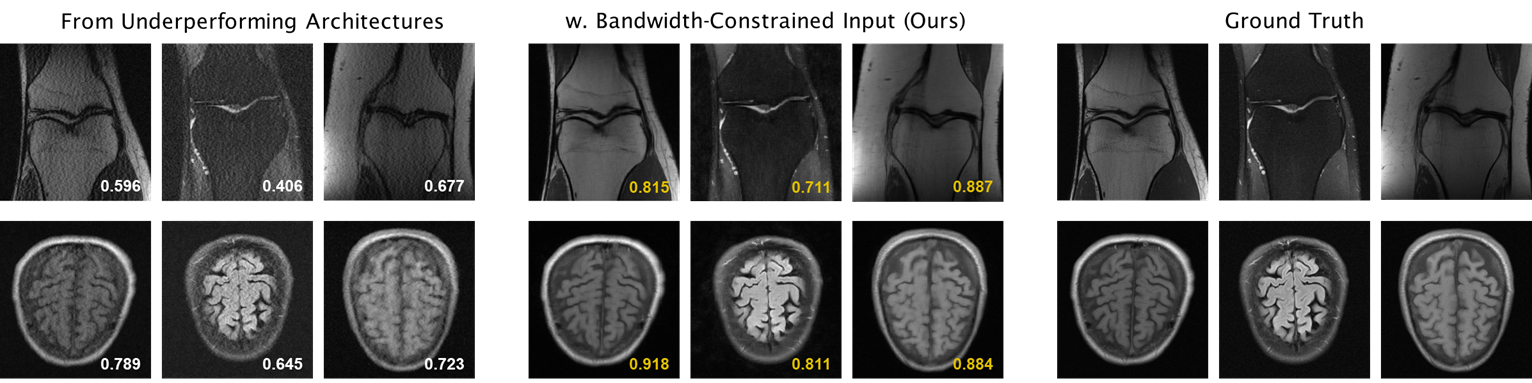}
    \caption{\textbf{Improving underperforming architectures} (SSIM ($\uparrow$)). Turning the left to the right simply by replacing the white-noise input with \textit{selected} Fourier features or low-pass filtering the noise input, which can be implemented with \textit{a few lines of code}.} 
    \label{fig:1}
\end{figure}

Instead of requiring large-scale datasets for capturing prior statistics, untrained networks \cite{heckel2018deep} inspired by deep image prior (DIP) \cite{ulyanov2018deep} use only the corrupted or partial measurements and a task-specific forward operator. Reconstruction is regularized solely by the inductive biases of the network architectures, enabling zero-shot self-supervised reconstruction for various imaging inverse problems \cite{yu2022validation,liu2023deepeit,qayyum2022untrained}. Concretely, a CNN, which parameterizes the unknown desired image, is optimized such that the output image, transformed by the forward operator, matches the acquired measurements. Such parameterization exhibits surprisingly high resistance to noise and corruption, which acts as a form of implicit regularization. Recent studies have attributed this property to CNN's inherent spectral bias---the tendency to fit the low-frequency signals before the high-frequency signals (e.g., noise) \cite{shi2022measuring,chakrabarty2019spectral}. The choice of network architecture is shown to be critically relevant to such bias \cite{chakrabarty2019spectral,liu2023devil,arican2022isnas}. 

Despite the great promise, obtaining favorable results with untrained network priors is contingent upon two critical factors: an optimal architecture specific to a task and an early-stopping strategy to prevent overfitting to noisy or partial measurements. Furthermore, optimizing on a per-image basis makes the reconstruction process domain-agnostic but inherently slow \cite{yaman2021zero}. While these issues are intertwined, with overfitting and runtime issues exacerbated by inappropriate and over-parameterized architectures, most existing efforts tackle these challenges separately. For architectural design, existing methods rely on either handcrafting or utilizing neural architecture search techniques \cite{heckel2018deep,darestani2021accelerated, arican2022isnas,chen2020dip,korkmaz2022unsupervised,ho2021neural}. However, the lack of consensus on architectural priors often leads to laborious search. Another line of work is dedicated to preventing overfitting through oracle early-stopping criterions \cite{yaman2021zero,wang2021early}, subspace optimization \cite{barbano2023fast} or pretraining then fine-tuning \cite{barbano2022educated,nittscher2023svd}. These methods mostly use fewer trainable parameters or hold out a subset of measurements for self-validation, in the spirit of traditional strategies, and often trade-off accuracy or involve costly pre-training.

In this work, we explore the possibility of modulating the frequency bias and hence the regularization effects of network priors in an \textit{architecture-agnostic manner}, aiming to enhance the performance of a given architecture irrespective of its configuration specifics. This is conceivable in light of the recent body of theoretical and empirical evidence indicating that there are \textit{only a few} key components (e.g., unlearnt upsampling) within the architecture that are the driving forces behind the spectral bias in DIP \cite{liu2023devil,chakrabarty2019spectral,shi2022measuring,heckel2020denoising}. Motivated by these findings, we develop efficient methods from a frequency perspective to effectively regularize the network priors, alleviating overfitting by curbing the overly rapid convergence of undesired high-frequency components, all with minimal architectural modifications and computational costs. Specifically, we propose to (1) constrain the effective bandwidth of the input via blurring or using Fourier features with a \textit{narrower} frequency range, (2) adjust the bandwidths of the interpolation-based upsamplers with controllable attenuation (smoothing) extents, and (3) regularize the Lipschitz constants of the convolutional layers to enforce function smoothness. 

We found empirically that by mitigating convergence to high-frequency components, our regularized network priors not only exhibit less vulnerability to overfitting but also tend to achieve better extrapolation capabilities in inpainting tasks. In the context of MRI reconstruction, which is essentially an inpainting task occurring in $k$-space, our methods significantly improve models across various architectural configurations without necessitating extensive architectural tuning (\cref{fig:1}). Their efficacy is also showcased in denoising and inpainting for natural images. By minimizing architectural influences, our approach additionally offers a unique advantage in efficiency: a \textbf{smaller}, previously underperforming network, can now achieve performance on-par with or even surpasses a \textbf{larger}, heavily parameterized high-performing network. Our contribution is three-fold: 
    \begin{itemize}
    \item We propose efficient techniques to directly modulate the frequency bias in untrained network priors, addressing architectural design, overfitting, and runtime challenges in a unified, architecture-agnostic manner.
    
    \item 
    The enhanced untrained networks match leading self-supervised methods with up to 90$\times$ faster runtime (1\,hr/slice to $\sim$5 \,min/slice) for MRI reconstruction and surpass supervised methods on out-of-domain data.
    
    \item Our findings on medical and natural image reconstruction reveal the spectral behaviors of CNNs in a single-instance generative setting.

\end{itemize}

\section{Related Work}
\textbf{Spectral Bias and function smoothness.} Function smoothness, also referred to as function frequency, quantifies how much the output of a function varies with changes in its input \cite{fridovich2022spectral}. Spectral bias \cite{rahaman2019spectral,xu2019training} is an implicit bias that favors learning functions changing at a slow rate (low-frequency), e.g., functions with a small Lipschitz constant. In visual domains, this is evident from the lack of subtle details in network outputs. Many regularization techniques to aid generalization, such as early stopping and $\ell_2$ regularization \cite{rosca2020case,nakkiran2021deep}, implicitly encourage smoothness. To explicitly promote smoothness, it is natural to penalize the norm of the input-output Jacobian \cite{novak2018sensitivity,hoffman2019robust}. However, computation of the Jacobian matrix for the high-dimensional MRI data during training is very expensive. Another efficient and prevalent solution is to constrain the network to be \textit{c}-Lipschitz with a \textit{pre-defined} Lipschitz constant \textit{c} \cite{miyato2018spectral,gouk2021regularisation}. We followed this line with a novel aim of achieving architecture-agnostic untrained image reconstruction. 

\textbf{Input frequency and overfitting.} The network input plays an important role in helping the neural network represent signals of various frequencies. Pioneering work on spectral bias \cite{rahaman2019spectral} showed theoretically and empirically that fitting high frequencies becomes easier, provided that the data manifold itself contains high-frequency components (Sec.~4 in \cite{rahaman2019spectral}). This has directly motivated implicit neural representations (INRs) \cite{sitzmann2020implicit} and neural radiance fields (NeRFs) \cite{mildenhall2021nerf} where coordinates are mapped to RGB values: naively training with raw coordinates as inputs results in over-smoothing; encoding the input coordinates with sinusoidal functions of higher frequencies enables the network to represent higher frequencies \cite{mildenhall2021nerf,tancik2020fourier}. However, it has recently been reported that the high-frequency bands of NeRF's input encodings incur overfitting and lead to failure in few-shot settings \cite{yang2023freenerf}. We found similar issues in untrained networks and propose to constrain the input's frequency range to counteract the convergence of high frequencies. 

\textbf{Architecture-induced spectral bias.} Recent studies on the working mechanisms of DIP reveal that unlearnt upsampling, with the low-pass characteristics of its interpolation filter, is responsible for the regularizing effects of DIP \cite{liu2023devil,heckel2020denoising,chakrabarty2019spectral}. Liu et al. \cite{liu2023devil} showed that the fixed upsampling operations readily bias the architecture towards producing low-frequency outputs, critically influencing both the peak PSNR and the starting point at which performance decays. The convolutional layer is another core element that exhibits stronger frequency selectivity compared to fully-connected layers and 1D layers \cite{chakrabarty2019spectral,shi2022measuring}. These findings motivated us to operate directly on these elements to achieve architecture-agnostic control.

\textbf{Avoiding overfitting} is a primary goal of unsupervised reconstruction where only noisy or partial measurements are available. Wang et al. \cite{wang2021early} proposed an early-stopping (ES) criterion by tracking the running variance of the output, but it is found to be unstable in medical settings \cite{barbano2023fast}. Yaman et al. \cite{yaman2021zero} proposed to split the available measurements into training and validation subsets and use the latter for self-validation for automated early stopping. While ES prevents overfitting, it cannot enhance the capability of a given architecture like our approach. Transfer-learning based methods aim to use fewer non-trainable parameters by performing pre-training followed by fine-tuning \cite{barbano2022educated,nittscher2023svd} or subspace optimization \cite{barbano2023fast}. In contrast, our method directly modulates the spectral bias to mitigate the convergence of undesired high frequencies, which is found to also improve the networks' extrapolation capabilities, all while maintaining the model complexity. 

\begin{figure}[t!]
    \centering
    \includegraphics[width=0.90\linewidth]{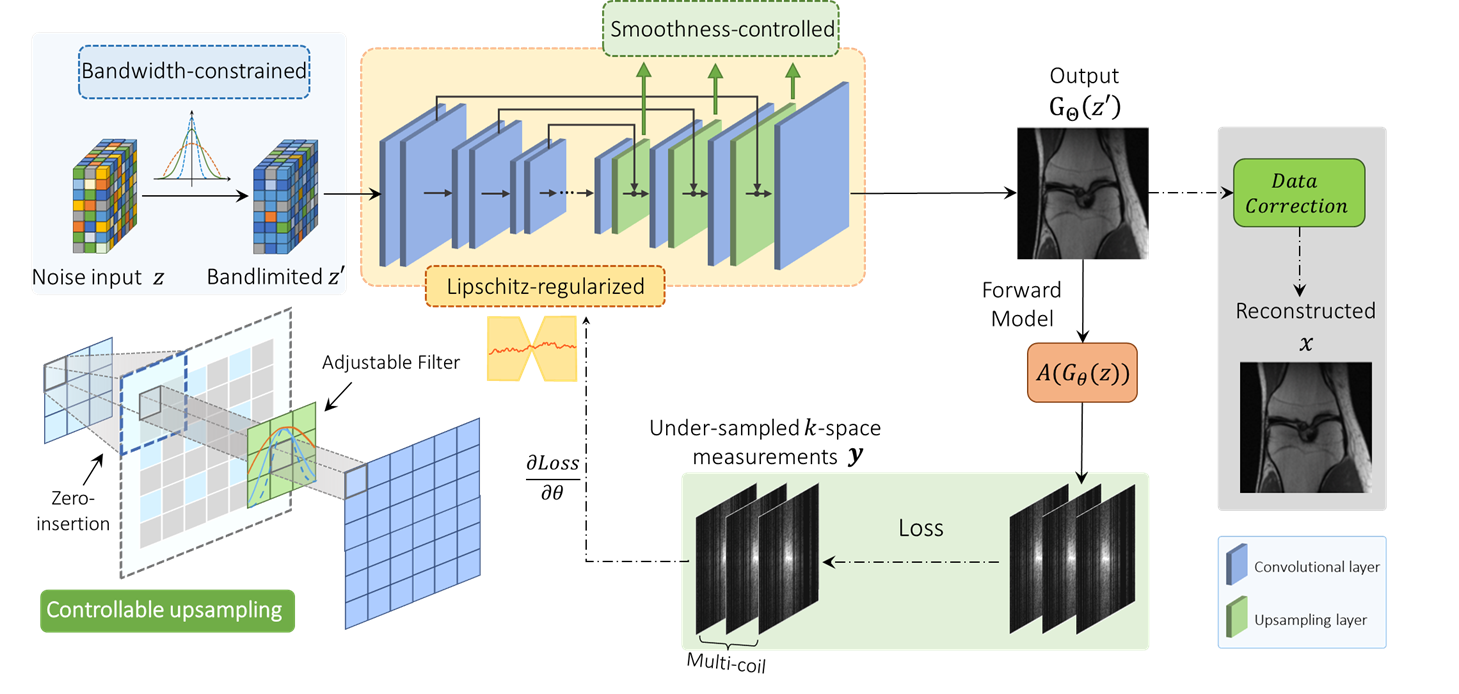}
    \caption{Overview of the proposed regularized network priors for MRI reconstruction. By adjusting the bandwidth of the input, the interpolation-based upsampling, and regularizing the convolutional layers, our approach enables more direct control over the spectral bias of architectures with various depths and widths.  } \label{fig:overview}
\end{figure}

\section{Method}
\subsection{Preliminaries}
\textbf{Accelerated MRI} The goal of accelerated MRI reconstruction is to recover a desired image \(\mathbf{x \in \mathbb{C}}^n\) (\(n = n_{\text{h}}\times n_{\text{w}}\)) from a set of under-sampled \textit{k}-space measurements. We focus on a multi-coil scheme in which the forward model is defined as 
\begin{equation}
    \label{eq:1}
    \mathbf{y}_i =\mathbf{A}_i\mathbf{x} + \epsilon, \quad \mathbf{A}_i = \mathbf{MFS}_i, \quad i=1,\dots, c,
\end{equation}
where \(\mathbf{y}_i\in\mathbb{C}^{m}\) denotes the \(k\)-space measurements from coil \(i\), \(c\) denotes the number of coils, \(\mathbf{S}_i\in\mathbb{C}^{n}\) denotes the coil sensitivity map (CSM) that is applied to the image \(\mathbf{x}\) through element-wise multiplications, \(\mathbf{F}\in\mathbb{C}^{n\times n}\) denotes the 2D discrete Fourier transform, \(\mathbf{M}\in\mathbb{C}^{m\times n}\) denotes the under-sampling mask, and \(\epsilon\in\mathbb{C}^{m}\) denotes the measurement noise. 

\textbf{Untrained MRI Reconstruction} is often framed as an inpainting problem where the network recovers the unacquired $k$-space measurements (masked) based on the acquired $k$-space data (observed). The image \(\mathbf{x}\) is parameterized via a neural network \(\mathbf{G_{\theta}(z)}\) with a fixed noise input vector \(\mathbf{z}\) drawn from a uniform distribution $z\sim\mathcal{U}(0,1)$. With the MRI forward model in Eq.~\ref{eq:1}, the untrained network solves the following optimization problem: 
\begin{equation}
    \label{eq:objective}
    \theta^\ast = \mathop{\arg \min}\limits_{\theta}\mathcal{L}(\mathbf{y};\mathbf{AG_{\theta}(z)}), \quad \mathbf{x}^\ast = \mathbf{G_{\theta^{\ast}}(z)}.
\end{equation} 
This parameterization enables the design of novel image priors based on the network architecture and its parameters, rather than on handcrafted image space priors. Nevertheless, many studies augment the untrained networks with traditional image regularizers \cite{liu2019image}, e.g., total variation (TV), which promotes piecewise constant images and can only partially alleviate over-fitting \cite{nittscher2023svd,barbano2023fast}. 

\subsection{Architecture-Agnostic Frequency Regularization}
To modulate the regularization effects of the network prior, we identify three core elements within the framework of DIP that lead to spectral bias and introduce corresponding regulation methods, as depicted in \cref{fig:overview}. 

\begin{figure}[t!]
\centering
\includegraphics[width=0.95\textwidth]{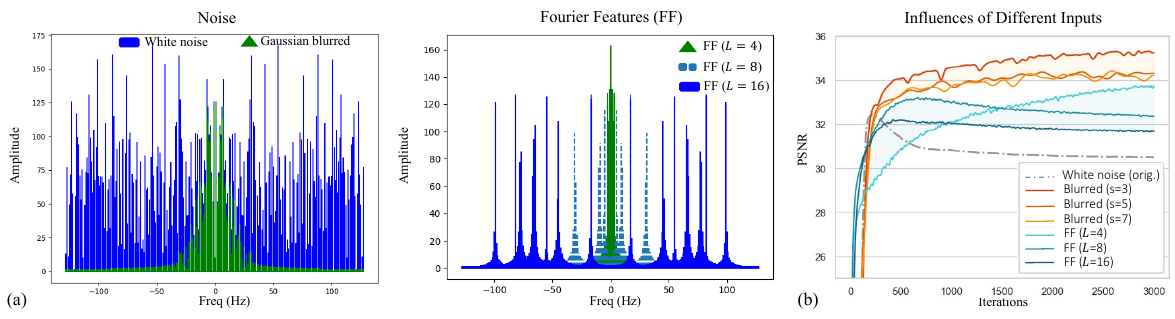} 
\caption{(a) Visualization of the 1D white noise, the low-passed noise via Gaussian blur, and Fourier features of various frequencies in the frequency domain. (b) Limiting the input's bandwidth via either Gaussian blur or Fourier features with lower $f_c$ ($L=4$ or $8$) are both effective in alleviating overfitting and enhancing the peak performance.    
} \label{fig:noise_fft}
\end{figure}

\textbf{Bandwidth-Constrained Input.} Inspired by implicit neural representations (INRs), we rethink the role of inputs in untrained networks for learning different frequencies. Conventionally, the inputs are
randomly sampled from either a uniform or Gaussian distribution and are then
mapped to image intensities.  From a frequency perspective, such input comprises all frequencies with uniform intensities \cite{e2008digital}, as white noise with variance $\sigma^2$ exhibits an autocorrelation that is a scaled Dirac $\delta$-function $\sigma^2\delta(t)$, whose Fourier transform $\mathcal{F}$ has a constant magnitude $\sigma^2$ spanning all frequencies $\mu$, i.e., $\mathcal{F}\{\sigma^2\delta(t)\}(\mu)=\sigma^2$, $\mu\in\mathbb{R}$. With this view in mind, we draw an analogy between untrained networks and INRs that map Fourier features to RGB values. Fourier features are sinusoid functions of the input coordinates $\mathrm{\mathbf{p}}$, i.e., $[\sin(\mathrm{2^0\pi\mathbf{p}}),\cos(\mathrm{2^0\pi\mathbf{p}}),...,\sin(2^{L-1}\pi\mathbf{\mathrm{\mathbf{p}}}),\cos(2^{L-1}\pi\mathbf{\mathrm{\mathbf{p}}})]$, where a larger $L$ assists the network in representing higher frequencies \cite{tancik2020fourier}. This is consistent with \cite{rahaman2019spectral} which indicates that the frequency magnitudes that can be expressed by the network increase with increasing frequency in the data manifold.  

In this sense, an untrained network can be viewed as \textit{mapping a broad spectrum of Fourier features to a target image} (\cref{fig:noise_fft} (a)). We hypothesize that this promotes rapid convergence of high frequencies, which is likely to bias the network towards high-frequency artifacts and hinder its ability to exploit 
\begin{wrapfigure}{r}{0.5\textwidth}
\centering
\includegraphics[width=0.38\textwidth]{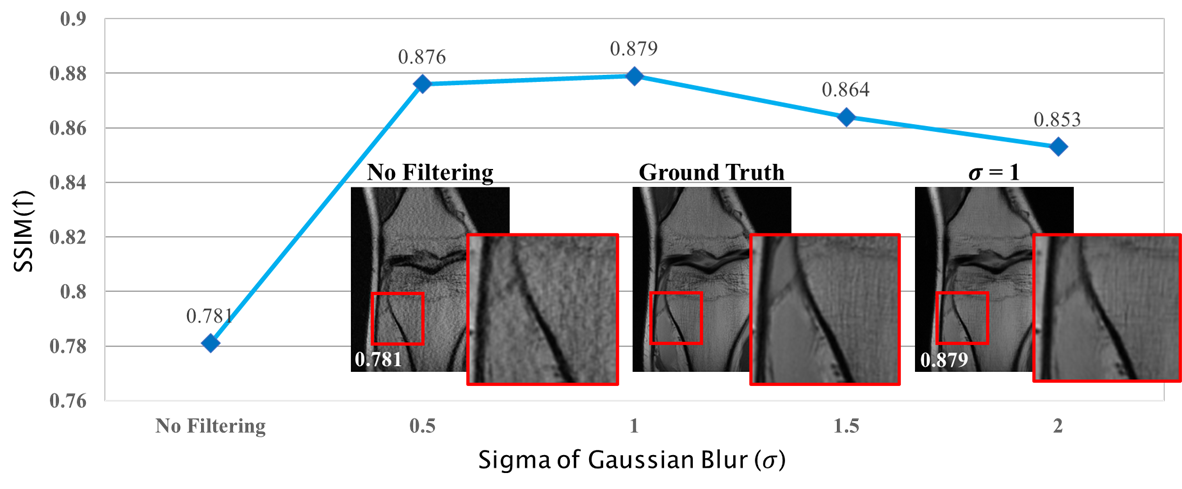} 
\caption{Narrowing input's bandwidth promotes low frequencies and enhances extrapolation capabilities as a "free lunch". The output becomes smoother as $\sigma$ increases, up to a certain point.
} \label{fig:4}
\end{wrapfigure}
spatial information effectively for extrapolation.

To validate this, we applied a Gaussian blur filter $\mathcal{G}_{s,\sigma}$ on the noise input $z$ to remove a certain amount of high frequencies before passing it to the network: $z*\mathcal{G}_{s,\sigma}$, where $*$ denotes convolution. The filter size $s$ and sigma value $\sigma$ that controls the filter's bandwidth are hyperparameters. As exemplified in Fig.~\ref{fig:4}, simply adjusting $\sigma$ already brings significant gains without architectural changes. Alternatively, as shown in \cref{fig:noise_fft} (b), substituting the noise input with Fourier features, with a carefully selected maximum frequency $f_c\propto L$ (e.g., $L=4$ or 8) to narrow the input's effective bandwidth, is also effective. As $L$ increases, e.g., $L=16$, the frequency range of Fourier features input approximates that of the original noise input, and the performance deteriorates similarly, further supporting our hypothesis. Fourier features introduce frequency-diverse input akin to white noise but in a \textit{controlled} manner, enabling regularization over the frequency selectivity of the network.

In light of this empirical evidence, we propose to \textit{limit} the input's bandwidth to mitigate the fitting of high-frequency components in untrained networks, achievable efficiently via blurring or Fourier features with a narrowed frequency range. Gaussian blur and Fourier Features offer flexible bandwidth control over the input through hyperparameters $\{s, \sigma\}$ and $L$, respectively, which allows for scaling under higher noise level/undersampling rates (examples of $8\times$ undersampling in Fig.~\ref{fig:8times}). We experimented with more sophisticated schedulers for the hyperparameters, but found that simply fixing them throughout the training yields superior performance.  




\begin{table}[b!]
\caption{\textbf{Influences of upsamplers on reconstruction}. From the left to the right, the attenuation extent of the upsampling method increases. Construction details of the upsampler $\mathcal{L}_{-90}$ follows \cite{liu2023devil}. Frequency responses of the interpolation filters are shown in the figure below. Evaluated on the $4\times$ fastMRI multi-coil brain datasets.} \label{ta:upsampling_mri}
\centering
\begin{adjustbox}{width=0.8\linewidth,center}
\begin{tabular}{@{}ccccc|c@{}}
\toprule
Methods      & w/o. Upsampling. & Nearest    & Bilinear & $\mathcal{L}_{-90}$ & \# of Params. (Millions) \\ \midrule
ConvDecoder \cite{darestani2021accelerated}  & 28.69 $\pm$ 1.6            & 31.78 $\pm$ 1.2  &  32.31 $\pm$ 1.3        & 32.48 $\pm$ 1.2                       & 4.1 M                                        \\
Deep Decoder \cite{heckel2018deep} & 24.55 $\pm$ 1.1             & 27.10 $\pm$ 0.9 & 31.36 $\pm$ 1.4    & 32.68 $\pm$ 1.1                        & 0.47 M                                       \\ \bottomrule
\end{tabular}
\end{adjustbox}

\end{table}

\begin{figure}[b!]
    \centering
    \includegraphics[width=0.75\linewidth]{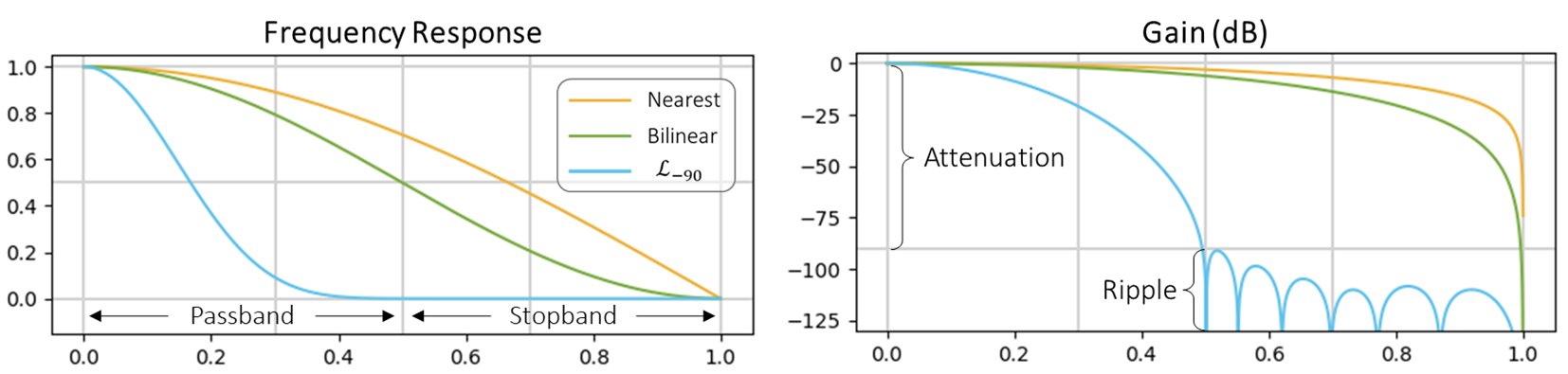}
\end{figure}

\textbf{Bandwidth-Controllable Upsampling.} We observed that only constraining the input's bandwidth significantly enhances \textit{shallower} architectures, yet the improvement diminishes as the network depth increases (Tab.~\ref{ta:ablation_brain},~\ref{ta:ablation_knee}). This could be partly attributed to the increased network layers that can always generate new arbitrarily high frequencies \cite{rahaman2019spectral,karras2021alias}. 

We note that the interpolation-based upsampling methods within the network, such as nearest neighbor and bilinear, essentially act as implicit low-pass filters, attenuating the alias frequencies caused by the increased sampling rate of the input feature maps. Prior works \cite{liu2023devil,heckel2020denoising} in denoising have shown that these non-trainable upsampling methods are driving forces behind the spectral bias of DIP, delaying the convergence of higher frequencies. Different upsamplers bias the network towards different spectral properties, depending on the bandwidth of the interpolation filter \cite{liu2023devil}. Here, we show in \cref{ta:upsampling_mri} that upsampling also substantially influences the performance image reconstruction.     

Motivated by these results, we introduce an upsampler with \textit{controllable bandwidth} so as to modulate the network's spectral bias, especially for \textit{deeper} architectures. We construct it by 1) first interleaving the input feature maps with zeros, and then 2) convolving them with a customized low-pass filter with adjustable bandwidth (\cref{fig:overview}). For filter design, we adopt the Kaiser-Bessel window \cite{kaiser1974nonrecursive} as it offers explicit control over the tradeoffs between passband ripple and stopband attenuation. The Kaiser window is defined as 
\begin{equation}
w(n) = I_0 (\beta\sqrt{1-(2n/M)^2})/I_0(\beta), -M/2\leq n\leq M/2,
\end{equation} where $M$ is the desired spatial extent of the window, $\beta\geq0$ is the shape parameter---the higher it is, the greater the stopband attenuation is (and generally the smoother the image is), and $I_0$ is the zeroth-order modified Bessel function of the first kind. This plug-and-play upsampler can be \textit{inserted in different layers with different $M$ and $\beta$ hyperparameters}, offering flexible and precise control.

\textbf{Lipschitz-Regularized Layers.} Compared to the non-trainable upsampling that only attenuates signals, the network layer with nonlinearies is the \textit{only} operation capable of \textit{generating new frequencies} \cite{karras2021alias}. We regularize their Lipschitz constants to control their sensitivity to input variations, which in turn can affect the spectral bias. Formally, a function \(f: \mathcal{X}\rightarrow\mathcal{Y}\) is said to be Lipschitz continuous if there is a constant \(k>0\) such that
    $\|f(\mathrm{x_1})-f(\mathrm{x_2})\|_\text{p} \leq k \|\mathrm{x_1}-\mathrm{x_2}\|_\text{p} \quad   \forall\mathrm{x_1},\mathrm{x_2}\in\mathcal{X},$
where $k$ is the Lipschitz constant that bounds how fast $f$ can change globally w.r.t. input perturbations. Spectral bias towards low frequencies favors functions with small Lipchitz constants.

Instead of upper bounding the Lipschitz constants of the network layers to pre-defined and manually chosen values as in \cite{shi2022measuring}, we make the per-layer Lipschitz bounds learnable and regularize their magnitudes during optimization. 

The Lipschitz constant of a convolutional layer is bounded by the operator norm of its weight matrix \cite{gouk2021regularisation}. To bound a convolutional layer to a specific Lipschitz constant $k$, the layer with \(m\) input channels, \(c\) output channels and kernels of size \(w\times h\) is first reshaped to a 2-D matrix \(W^{\ell}\in\mathbb{R}^{n\times cwh}\), and normalized as
\begin{equation}
    \Tilde{W_{\ell}}=\frac{W_{\ell}}{\text{max}(1, \frac{\|W_\ell\|_\text{p}}{\mathtt{SoftPlus}(k_\ell)})},
\end{equation}
where \(k_\ell\) is a learnable Lipschitz constant for each layer, \(\|\cdot\|_\text{p}\) is chosen as the \(\ell_\infty\) norm  and \(\mathtt{SoftPlus}(k_\ell)=\text{ln}(1+\text{exp}(k_\ell))\) ensures the learned Lipschitz bounds are non-negative. Such formulation \textit{only} normalizes \(W_\ell\) if its matrix norm is larger than the learned Lipschitz constraint during training. Integrating the ultimate Lipschitz regularization into \cref{eq:objective}, our regularized training objective is
\begin{equation}
    \min_{\Theta,K} \mathcal{L}(\mathbf{y};\mathbf{AG_{\Theta}(z)}) + \lambda\sum_{l=1}^L \mathtt{SoftPlus}(\mathbf{k}_\ell)^2
\end{equation} where \(\mathbf{K}\) is a collection of per-layer learnable Lipschitz constant \(k_\ell\) jointly optimized with the network parameters, and \(\lambda\) controls the granularity of smoothness. 

\begin{figure}[t!]
    \centering
    \includegraphics[width=\linewidth]{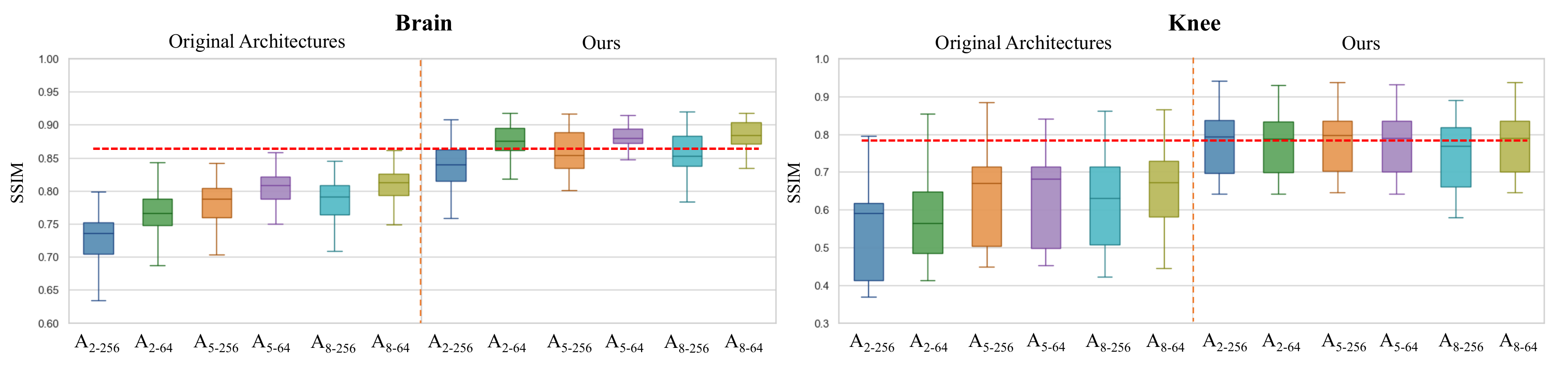}
    \caption{Our approach significantly minimizes the performance gaps among architectures with various depths $\{2,5,8\}$ and widths $\{64, 256\}$.}
    \label{fig:boxplot}
\end{figure}

\section{Experiments}
\subsection{Setup and Datasets}
We first (1) validate the effectiveness of the proposed methods in enhancing the performance of untrained networks across various architectural configurations, especially those originally underperforming. We then (2) benchmark the enhanced versions of those \textit{compact} architectures against established supervised and self-supervised methods on both in-domain and out-of-domain datasets in terms of accuracy and efficiency. We also (3) compare our methods with self-validation-based early stopping \cite{yaman2021zero} on overcoming overfitting, and show that it is complementary to our approach by further shortening the reconstruction time. Finally, we demonstrate the utility of our methods in (4) general image inpainting and denoising tasks and (5) perform spectral bias analysis on all evaluated tasks. 

The MRI experiments were performed on two publicly available datasets: the multi-coil knee and brain MRI images from \underline{fastMRI} database \cite{knoll2020fastmri}, and multi-coil knee MRI images from \underline{Stanford 3D FSE knee} dataset \cite{fsecreation}. The fully-sampled \(k\)-space data was retrospectively masked by selecting 25 central \(k\)-space lines along with a uniform undersampling at outer \(k\)-space, achieving the standard \(4\times\) acceleration. For training the supervised baseline, the knee training set consists of 367 PD and PDFS slices and the brain training set consists of 651 slices with a mixture of T1 and T2 weighted images. 50 knee slices and 50 brain slices were sampled from the respective multi-coil validation datasets for evaluation.

\subsection{Implementation Details} \label{sec:details}
Without loss of generality, the base architectures considered in our work are N-level encoder-decoder architectures with full skip connections. The architectures are isotropic with the same width and kernel size throughout the layers. All evaluated architectures are trained for 3k iterations using mean absolute error and Adam optimizer \cite{kingma2014adam} with a learning rate of 0.008. Unless otherwise specified, the results at the last iteration are reported. The input is drawn from a uniform distribution $z\sim\mathcal{U}(0,1)$. The filter size of the Gaussian blur was set to 5 and the sigma value was randomly sampled from $[0.5, 2.0]$ for every slice. $M$ and $\beta$ for the Kaiser-based upsamplers are chosen to be $\{15\times N-1, 5\}$ and $\{5\times N\}$ for knee data ($N$ denotes the nth-level), and $\{5\times N\}$ and $\{5\times N\}$ for brain data, respectively. $\lambda$ is set to \(1\) for the Lipschitz regularizer. 

\begin{table}[t!]
\caption{\textbf{Effectiveness of the methods in bridging performance gaps among different architectures}, evaluated on fastMRI \textbf{brain} datasets. Bandlimited inputs achieved by Fourier features ($L=4$ or $8$) or Gaussian blur along with Lipschitz regulariation improve all architectures, especially the shallower. The proposed Kaiser-based upsampling dramatically improves the deeper architectures. All architectures end up with similarly high performance. The \colorbox{pink!50}{best} and the \colorbox{blue!15}{second-best} are highlighted. } \label{ta:ablation_brain}
\centering
\begin{adjustbox}{width=0.95\columnwidth,center}
\begin{tabular}{@{}l|cccccc|cccccc@{}}
\toprule
Regularizers            & \multicolumn{1}{l}{$\mathbf{A_2\mathunderscore_{256}}$} & \multicolumn{1}{l}{$\mathbf{A_2\mathunderscore_{64}}$} & \multicolumn{1}{l}{$\mathbf{A_5\mathunderscore_{256}}$} & \multicolumn{1}{l}{$\mathbf{A_5\mathunderscore_{64}}$} & \multicolumn{1}{l}{$\mathbf{A_8\mathunderscore_{256}}$} & \multicolumn{1}{l|}{$\mathbf{A_8\mathunderscore_{64}}$} & \multicolumn{1}{l}{$\mathbf{A_2\mathunderscore_{256}}$} & \multicolumn{1}{l}{$\mathbf{A_2\mathunderscore_{64}}$} & \multicolumn{1}{l}{$\mathbf{A_5\mathunderscore_{256}}$} & \multicolumn{1}{l}{$\mathbf{A_5\mathunderscore_{64}}$} & \multicolumn{1}{l}{$\mathbf{A_8\mathunderscore_{256}}$} & \multicolumn{1}{l}{$\mathbf{A_8\mathunderscore_{64}}$} \\ \midrule
                     & \multicolumn{6}{c|}{PSNR \(\uparrow\)}                                                                                                                                                                         & \multicolumn{6}{c}{SSIM \(\uparrow\)}                                                                                                                                                                         \\ \cmidrule{2-13}
w/o. Reg. (Plain) & 29.08                          & 29.41                         & 31.15                          & 31.42                         & 31.27                         & 31.68                         & 0.729                         & 0.761                         & 0.782                          & 0.801                         & 0.784                          & 0.807                         \\
TV                   & 29.22                         & 29.61                         & 31.26                          & 31.37                         & 31.32                          & 31.64                          & 0.735                          & 0.764                         & 0.785                          & 0.802                         & 0.787                          & 0.807                         \\

Lipschitz Reg.          & 30.92                          & 29.73                         & 31.47                          & 32.11                         & 31.50                          & 32.03                          & 0.795                          & 0.766                         & 0.792                          & 0.812                         & 0.800                         & 0.820                         \\
Fourier features  ($L=16$)      & 30.57                         &  30.49                      & 31.57                          & 31.77                         &  31.77                         &  32.09                         & 0.786                          & 0.788                        & 0.794                         & 0.813                         & 0.799                         & 0.819                         \\
Fourier features  ($L=8$)      & 31.42                        & 31.98                       &  31.82                         & 32.42                         & 31.60                          &  32.45                         & 0.804                         & 0.833                        & 0.799                         & 0.831                         & 0.795                         & 0.834                         \\
Fourier features  ($L=4$)      &  31.92                       & 32.59                      &  31.87                          & 32.80                         & 31.71                          &  32.86                         & 0.840                        & 0.863                         &  0.799                        &  0.848                        & 0.793                        & 0.844                         \\

Gaussian blurred     & \cellcolor{pink!50}33.34                         & 32.67                        & \cellcolor{blue!15}32.14                          & 32.66                         & \cellcolor{blue!15}32.03                          & 32.92                          & \cellcolor{pink!50}0.870                          & 0.866                         & 0.811                          & 0.849                         & \cellcolor{blue!15}0.825                         & 0.849                         \\ 
Gauss. + Lips.      & \cellcolor{blue!15}32.90  & \cellcolor{pink!50}33.12 & 32.08 & \cellcolor{blue!15}32.83 & 31.70 & \cellcolor{blue!15}33.14 & \cellcolor{blue!15}0.855 & \cellcolor{blue!15}0.870 & \cellcolor{blue!15}0.815 & \cellcolor{blue!15}0.851 & 0.805 & \cellcolor{blue!15}0.849                       \\ 
Gauss. + Lips. + Kaiser Up.     & 32.50 & \cellcolor{blue!15}33.10  & \cellcolor{pink!50}33.00 & \cellcolor{pink!50}33.21  & \cellcolor{pink!50}33.09  & \cellcolor{pink!50}33.85  & 0.836  & \cellcolor{pink!50}0.874  & \cellcolor{pink!50}0.857   & \cellcolor{pink!50}0.876  & \cellcolor{pink!50}0.858  & \cellcolor{pink!50}0.885                      \\
   
\bottomrule
\end{tabular}
\end{adjustbox}
\end{table}
\begin{table}[t!]
\caption{Evaluated on fastMRI \textbf{knee} datasets. } \label{ta:ablation_knee}
\centering
\begin{adjustbox}{width=0.95\linewidth,center}
\begin{tabular}{@{}l|cccccc|cccccc@{}}
\toprule
Regularizers            & \multicolumn{1}{l}{$\mathbf{A_2\mathunderscore_{256}}$} & \multicolumn{1}{l}{$\mathbf{A_2\mathunderscore_{64}}$} & \multicolumn{1}{l}{$\mathbf{A_5\mathunderscore_{256}}$} & \multicolumn{1}{l}{$\mathbf{A_5\mathunderscore_{64}}$} & \multicolumn{1}{l}{$\mathbf{A_8\mathunderscore_{256}}$} & \multicolumn{1}{l|}{$\mathbf{A_8\mathunderscore_{64}}$} & \multicolumn{1}{l}{$\mathbf{A_2\mathunderscore_{256}}$} & \multicolumn{1}{l}{$\mathbf{A_2\mathunderscore_{64}}$} & \multicolumn{1}{l}{$\mathbf{A_5\mathunderscore_{256}}$} & \multicolumn{1}{l}{$\mathbf{A_5\mathunderscore_{64}}$} & \multicolumn{1}{l}{$\mathbf{A_8\mathunderscore_{256}}$} & \multicolumn{1}{l}{$\mathbf{A_8\mathunderscore_{64}}$} \\ \midrule
                     & \multicolumn{6}{c|}{PSNR \(\uparrow\)}                                                                                                                                                                         & \multicolumn{6}{c}{SSIM \(\uparrow\)}                                                                                                                                                                         \\ \cmidrule{2-13}
w/o. Reg. (Plain) &  27.18                         & 27.62                        & 29.16                          & 29.23                        & 28.98                        &  29.35                       & 0.541                        &  0.575                        &  0.628                        &  0.640                       &  0.625                        &  0.644                       \\  
TV                   &   28.25                       &  27.85                       &   29.33                        &   29.57                       & 29.54                        &  30.01                         & 0.588                         & 0.592                        &  0.635                         & 0.651                        &  0.645                         &  0.687                      \\

Lipschitz Reg.        &  28.41                        & 29.21                         & 29.17                          &  29.79                        &  29.43                         & 30.14                          & 0.601                          &  0.600                      & 0.629                        & 0.651                       &  0.636                        & 0.666                         \\
Fourier features ($L=16$)      & 28.42                         & 28.97                       & 29.58                          & 30.26                         &  29.76                         & 30.38                          &  0.587                         & 0.622                         & 0.653                         & 0.671                         & 0.661                          & 0.681                         \\
Fourier features ($L=8$)      & 28.61                        & 29.98                       & 29.86                          &  30.72                        & 29.66                          & 30.89                          & 0.604                          & 0.670                         & 0.669                         &  0.693                        & 0.662                         & 0.703                          \\
Fourier features ($L=4$)      & \cellcolor{pink!50}32.02                         & \cellcolor{pink!50}32.07                       & 29.40                           &  31.13                        &  29.55                         & 31.17                          & \cellcolor{blue!15}0.775                          & \cellcolor{pink!50}0.781                         & 0.665                         & 0.718                         &   0.668                       &  0.717                        \\

Gaussian blurred     & 30.87                        &  30.89                       & \cellcolor{blue!15}30.02                         & 31.24                         &  29.31                         &  30.89                        & 0.739                           & 0.768                         & 0.694                          & \cellcolor{blue!15}0.748                        & \cellcolor{blue!15}0.698                       & 0.727                        \\
Gaussian blurred + Lips.    & 31.61  & \cellcolor{blue!15}31.93  & 29.40  & \cellcolor{pink!50}31.67  & \cellcolor{blue!15}29.82  & \cellcolor{blue!15}31.58  & 0.750  & \cellcolor{blue!15}0.776  & \cellcolor{blue!15}0.702  & 0.727  & 0.697  & \cellcolor{blue!15}0.732                      \\ 
Gauss. + Lips. + Kaiser Up.     & \cellcolor{blue!15}31.92 & 31.61  & \cellcolor{pink!50}31.78  & \cellcolor{blue!15}31.60 & \cellcolor{pink!50}31.09 & \cellcolor{pink!50}31.73 & \cellcolor{pink!50}0.777 & 0.776  & \cellcolor{pink!50}0.778  & \cellcolor{pink!50}0.776  & \cellcolor{pink!50}0.750  & \cellcolor{pink!50}0.768                      \\
\bottomrule
\end{tabular}
\end{adjustbox}
\end{table}

\subsection{Effectiveness in Reducing Architectural Sensitivity}
\begin{figure}[t!]
    \centering
    \includegraphics[width=\linewidth]{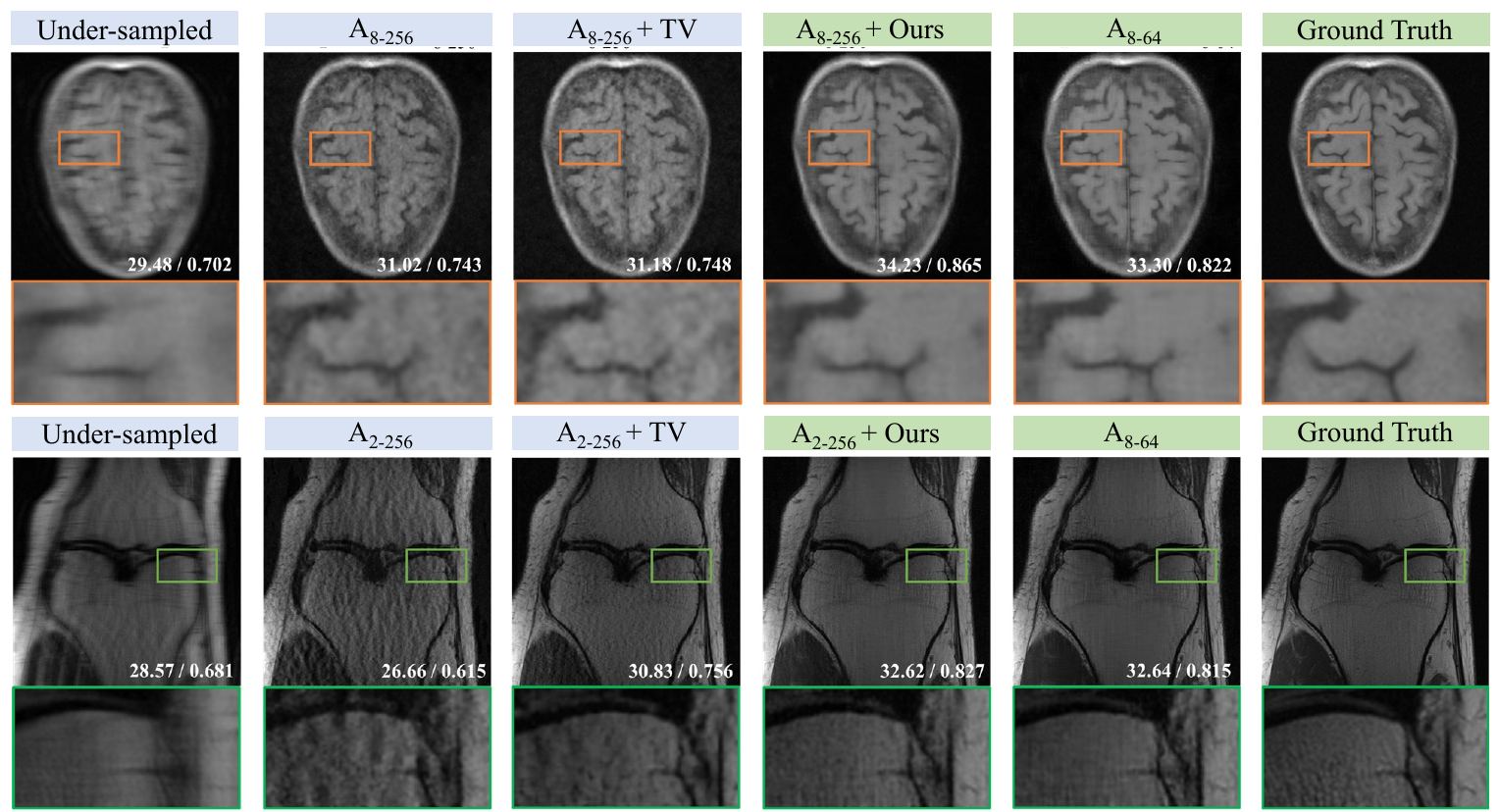}
    \caption{Our methods enable the underperforming architectures (e.g., $\mathbf{A_8\mathunderscore_{256}}$, $\mathbf{A_2\mathunderscore_{256}}$) to perform similarly to the well-performing architectures (e.g., $\mathbf{A_8\mathunderscore_{64}}$).}
    \label{fig:ablation_qua}
\end{figure}

\cref{fig:boxplot} gives a quantitative overview of the substantial improvement give by our approach in architectures with diverse configurations on both knee and brain datasets. The different results of the original architectures also confirm the influences of architectural choices on performance. Notably, before applying our methods, the deeper and narrower architectures tend to perform better than their counterparts (more in appendix). This trend aligns with previous works \cite{darestani2021accelerated,barbano2022educated,darestani2022test, ulyanov2018deep} where these architectures tend to be favored in inpainting-like tasks. Here we identify particularly their counterparts (i.e., $\mathbf{A_x\mathunderscore_{256}}$) as "underperforming" architectures. As will be shown in our spectral bias analysis in \cref{sec:freq_analysis} and appendix, these underperforming architectures learn high frequencies more quickly (though this may be desired for other tasks \cite{liu2023devil}) and are more susceptible to overfitting, incurring severe artifacts in the output (\cref{fig:ablation_qua,fig:inpainting_denoising}). When applied with our methods, as detailed in \cref{ta:ablation_brain} and \cref{ta:ablation_knee}, a large boost in performance is observed in all architectures, especially $\mathbf{A_2\mathunderscore_{256}}$. 

We observe that using low-passed inputs via either selected Fourier features or blurring brings the most benefits to the shallower architectures. Better results are achieved when combined with Lipschitz regularization on the layers. On the other hand, deeper architectures benefit more from the Kaiser-based upsampler, which can be seen as performing low-pass filtering on the input feature maps within the network, beyond the initial input layer. We further note that the hyperparameters required for upsampling differ between knee and brain data (\cref{sec:details}), with the knee data requiring greater attenuation. This is also consistent with previous findings that they require different numbers of channels for the architectural choices \cite{darestani2021accelerated}. Our methods greatly alleviate the need for such architectural tuning by instead allowing for the adjustment of a few key hyperparameters. 

\begin{table}[b]
\caption{Quantitative results on fastMRI datasets. Runtime: mean (std) per slice.} \label{ta:benchmark}
\centering
\begin{adjustbox}{width=0.82\linewidth,center}
\begin{tabular}{@{}llcccccccccc@{}}
\toprule
\multicolumn{1}{c}{Datasets} &                          & \begin{tabular}[c]{@{}c@{}}Supervised\\ UNet\end{tabular} & \begin{tabular}[c]{@{}c@{}}CS-$\ell_1$\\ \cite{jaspan2015compressed}\end{tabular} & \begin{tabular}[c]{@{}c@{}}ZS-SSL \\ \cite{yaman2021zero}\end{tabular}   & \begin{tabular}[c]{@{}c@{}}DIP \\ \cite{ulyanov2018deep}\end{tabular} & \begin{tabular}[c]{@{}c@{}}Deep Decoder\\ \cite{heckel2018deep}\end{tabular} & \begin{tabular}[c]{@{}c@{}}ConvDecoder\\ \cite{darestani2021accelerated}\end{tabular} & \begin{tabular}[c]{@{}c@{}}$\mathbf{A_2\mathunderscore_{64}}$\\ (vanilla)\end{tabular} & \begin{tabular}[c]{@{}c@{}}$\mathbf{A_8\mathunderscore_{64}}$\\ (vanilla)\end{tabular} & \begin{tabular}[c]{@{}c@{}}$\mathbf{A_2\mathunderscore_{64}}$\\ (Ours)\end{tabular} & \begin{tabular}[c]{@{}c@{}}$\mathbf{A_8\mathunderscore_{64}}$\\ (Ours)\end{tabular} \\ \midrule
\multirow{2}{*}{Brain}       & \multicolumn{1}{c}{PSNR $\uparrow$} & 33.35                                                     & 29.91 & \textbf{34.39}                                          & 31.15                                              & 26.97                                                   & 31.81                                                   & 29.42                                                & 31.68                                                & 33.10                                               & {\ul 33.85}                                         \\
                             & \multicolumn{1}{c}{SSIM $\uparrow$}                     & \textbf{0.889}                                            & 0.773 & {0.878}                                             & 0.782                                              & 0.747                                                   & 0.800                                                   & 0.761                                                & 0.807                                                & 0.874                                               & {\ul 0.885}                                \\ \midrule
\multirow{2}{*}{Knee}        & \multicolumn{1}{c}{PSNR $\uparrow$} & 31.15                                                     & 28.23 & \underline{32.00}                                          & 29.16                                              & 27.21                                                   & 29.59                                                   & 27.62                                                & 29.35                                                & \textbf{32.07}                                               & {31.73}                                         \\
                             & \multicolumn{1}{c}{SSIM $\uparrow$}                      & \underline{0.776}                                               & 0.633 & {0.773}                                             & 0.628                                              & 0.687                                                   & 0.655                                                   & 0.575                                                & 0.644                                                & \textbf{0.781}                                      & 0.768                                               \\ \midrule
\multicolumn{2}{l}{GFLOPS $\downarrow$}                              & 99.24                                                        & --    & \multicolumn{1}{c}{5461.6}                                    & \multicolumn{1}{c}{615.72}                               & \multicolumn{1}{c}{82.82}                                    & \multicolumn{1}{c}{699.94}                                    & 38.42                                                & \multicolumn{1}{c}{40.94}                                 & 62.36                                               & \multicolumn{1}{c}{68.38}                                \\
\multicolumn{2}{l}{Runtime (mins) $\downarrow$}                      & \begin{tabular}[c]{@{}c@{}}0.002 \\ (0.00003)\end{tabular}                                                        & --    & \begin{tabular}[c]{@{}c@{}}\textcolor{red}{64.8} \\ (20.18)\end{tabular} & \begin{tabular}[c]{@{}c@{}}14.0\\ (0.61)\end{tabular}    & \begin{tabular}[c]{@{}c@{}}6.6\\ (0.63)\end{tabular}         & \begin{tabular}[c]{@{}c@{}}8.2\\ (0.35)\end{tabular}         & \begin{tabular}[c]{@{}c@{}}5.4\\ (0.47)\end{tabular}      & \begin{tabular}[c]{@{}c@{}}10.5\\ (0.62)\end{tabular}      & \begin{tabular}[c]{@{}c@{}}\textcolor{cyan}{6.6}\\ (0.58)\end{tabular}     & \begin{tabular}[c]{@{}c@{}}12.3\\ (0.65)\end{tabular}     \\ \bottomrule
\end{tabular}
\end{adjustbox}
\end{table}

\begin{figure}[t!]
    \centering
    \includegraphics[width=\linewidth]{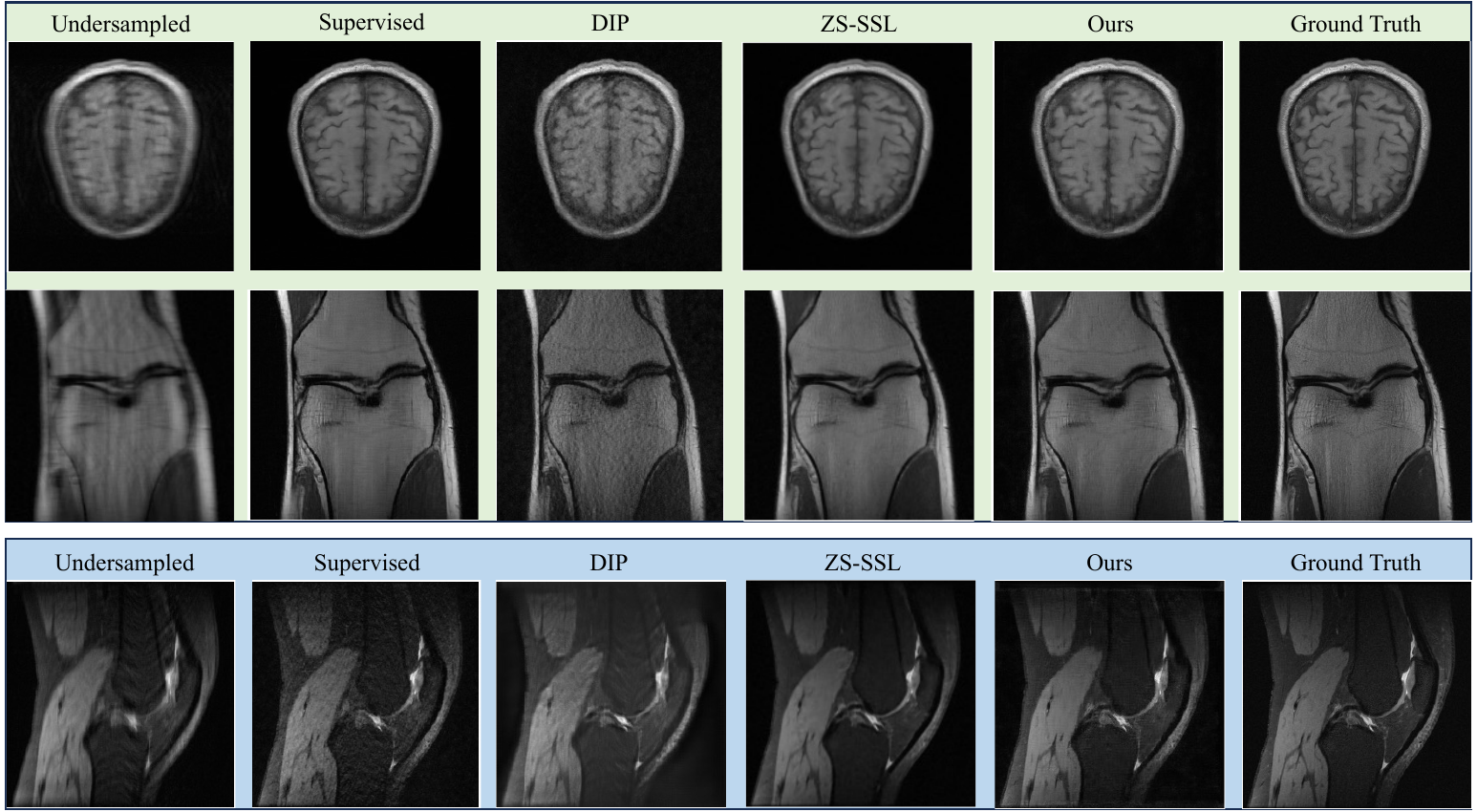}
    \caption{Qualitative evaluations. \colorbox{green!15}{In-domain}: fastMRI. \colorbox{blue!10}{Out-of-domain}: Stanford FSE.}
    \label{fig:qua_all}
\end{figure}

\begin{table}[b!]
\caption{\textbf{Out-of-domain evaluation} among supervised and untrained methods and comparisons with a \textbf{self-validation-based early stopping} strategy \cite{yaman2021zero}. } \label{ta:out_domain}
\centering
\begin{adjustbox}{width=0.7\linewidth,center}
\begin{tabular}{@{}lclccccc@{}}
\toprule
                                                & Method               & \multicolumn{2}{c}{\textbf{In-domain}}                         & \multicolumn{2}{c}{\textbf{Out-domain}}                         & \multicolumn{2}{c}{Runtime (mean$\pm$std)}               \\ \midrule
\multicolumn{1}{c|}{}  & \multicolumn{1}{l}{}  & \multicolumn{1}{c}{PSNR}                      & \multicolumn{1}{l}{SSIM}  & \multicolumn{1}{c}{PSNR}                      & \multicolumn{1}{c|}{SSIM}  & Train               & Inference                \\ \cmidrule(l){3-8} 
\multicolumn{1}{c|}{Trained}                           & U-Net                 & \multicolumn{1}{c}{31.16} & \underline{0.776}                     &  29.16                    & \multicolumn{1}{c|}{0.724} & $\sim$1.5 days & $0.1\pm0.003$  sec \\ \midrule
\multicolumn{1}{l|}{\multirow{5}{*}{Untrained}} &  CS-$\ell_1$ \cite{jaspan2015compressed} & \multicolumn{1}{c}{28.23} & 0.633                     & \multicolumn{1}{c}{22.46} & \multicolumn{1}{c|}{0.407} & --                  & --     \\ 
\multicolumn{1}{c|}{} 
& ZS-SSL \cite{yaman2021zero}  & \multicolumn{1}{c}{\underline{32.00}} & 0.773                     & \multicolumn{1}{c}{\textbf{31.74}} & \multicolumn{1}{c|}{\textbf{0.805}} & --                  & \textcolor{red}{$26.1\pm3.5$ mins}     \\

\multicolumn{1}{c|}{}
& DIP \cite{ulyanov2018deep}  & \multicolumn{1}{c}{29.16} & 0.628                     & \multicolumn{1}{c}{28.89} & \multicolumn{1}{c|}{0.664} & --                  & $9.2\pm 0.3$ mins      \\
\multicolumn{1}{c|}{}
& $\mathbf{A_2\mathunderscore_{64}}$  & \multicolumn{1}{c}{27.62} & 0.575                     & \multicolumn{1}{c}{26.03} & \multicolumn{1}{c|}{0.550} & --                  & $5.5\pm0.1$ mins     \\
\multicolumn{1}{c|}{}
& $\mathbf{A_2\mathunderscore_{64}}$  + Early Stopped  & \multicolumn{1}{c}{29.59} & 0.695                     & \multicolumn{1}{c}{27.59} & \multicolumn{1}{c|}{0.641} & --                  & $0.2\pm0.2$ mins     \\
\multicolumn{1}{l|}{}                           & $\mathbf{A_2\mathunderscore_{64}}$ (Ours)      & \multicolumn{1}{c}{\textbf{32.07}}                     & \multicolumn{1}{l}{\textbf{0.781}} & \underline{31.43}                     & \multicolumn{1}{c|}{0.790} & --                  & \textcolor{magenta}{$6.4\pm0.4$ mins}     \\
\multicolumn{1}{l|}{}                           & $\mathbf{A_2\mathunderscore_{64}}$ (Ours) + Early Stopped     & \multicolumn{1}{c}{31.97}                     & 0.776                     & 31.30                     & \multicolumn{1}{c|}{\underline{0.800}} & --                  & \textcolor{cyan}{$0.3\pm0.1$ mins}      \\ \bottomrule
\end{tabular}
\end{adjustbox}
\end{table}

\subsection{Benchmark Results}
We adopt two most lightweight architectures as the base models, and compare our regularized network priors with several established MRI reconstruction methods, including a supervised baseline, a state-of-the-art self-supervised method based on unrolling networks (\texttt{ZS-SSL}) \cite{yaman2021zero} and underparameterized untrained networks (\texttt{ConvDecoder} \cite{darestani2021accelerated}, \texttt{Deep Decoder} \cite{heckel2018deep}). Visual comparisons are presented in \cref{fig:qua_all}.  

\textbf{Comparisons with state-of-the-arts.} Our method enables the previously underperforming architectures to match the performance of \texttt{ZS-SSL} and that of the supervised UNet on fastMRI knee data (\cref{ta:benchmark}), and surpass the trained UNet on out-of-domain Stanford 3D FSE data (\cref{ta:out_domain}), demonstrating their advantages in generalizable reconstruction. Our enhanced networks also clearly outperform other lightweight untrained networks, i.e., \texttt{ConvDecoder} and \texttt{Deep Decoder}, which are designed to prevent overfitting.   $\texttt{ZS-SSL}$ is an unrolling method where the network, i.e., a ResNet \cite{timofte2017ntire}, is adopted as a denoiser. Our enhanced network priors achieve comparable performance than \texttt{ZS-SSL} while being orders of magnitude faster, thanks to the more efficient base models. 

\textbf{Comparisons with early stopping (ES).} $\texttt{ZS-SSL}$ uses a self-validation strategy for early stopping whereas our method does not necessitate ES. \setlength{\columnsep}{6pt}%
\begin{wrapfigure}{r}{0.3\linewidth}
\centering

\includegraphics[width=\linewidth]{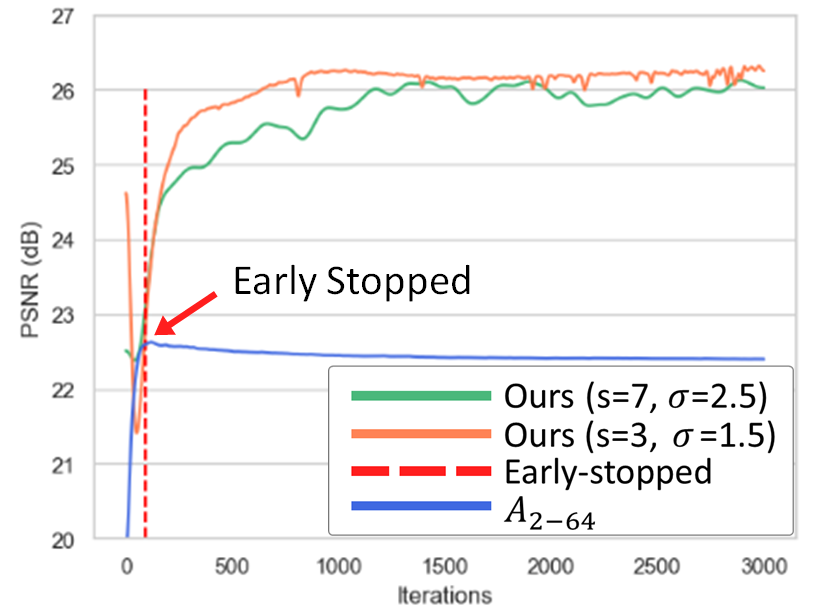}

\caption{$8\times$ undersampling}

 \label{fig:8times}
\end{wrapfigure} Our method not only alleviates overfitting but more importantly, enhances the inter/extrapolation capabilities of the underperforming architectures, leading to higher peak performance than the original scheme (Tab.~\ref{ta:out_domain}). This cannot be achieved even with the best ES strategy, as ES only halts the training near the peak PSNR for a given architecture, but it cannot fundamentally improve the underperforming architectures whose peak PSNR remains subpar (Fig.~\ref{fig:8times}). We show in Tab.~\ref{ta:out_domain} that self-validation based ES can be integrated into our approach to further shorten the reconstruction time from 6 mins$\rightarrow$0.3 mins. More examples are in Suppl.~Fig.~8.

\subsection{Spectral Bias Analysis} \label{sec:freq_analysis}
\begin{figure}[t!]
    \centering
    \includegraphics[width=0.92\linewidth]{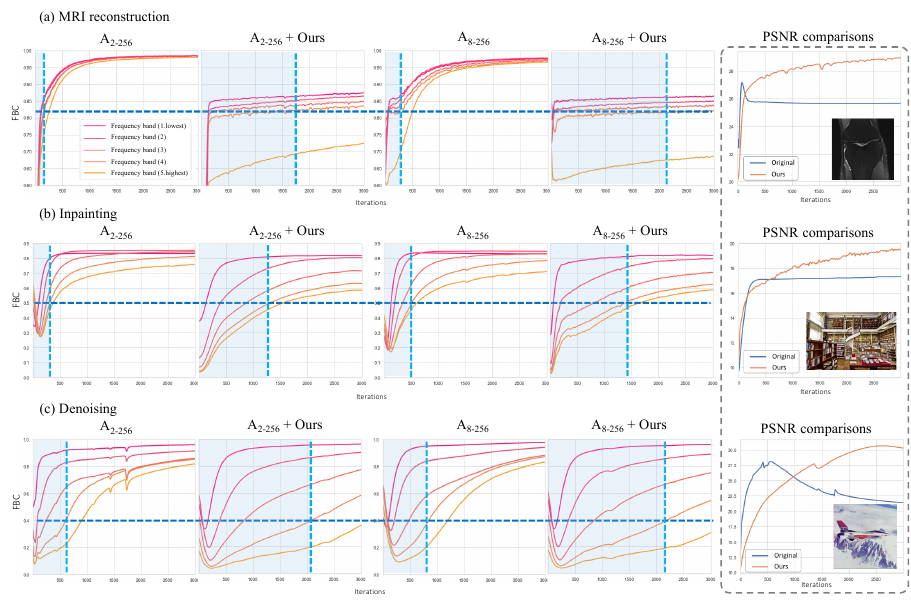}
    \caption{\textbf{Measurement of spectral bias}. Underperforming architectures (e.g., $A_2\mathunderscore_{256}$, $A_8\mathunderscore_{256}$) tend to learn high frequencies hastily, overfit more easily, and extrapolate poorly. Our method effectively mitigates these shortcomings.}
    \label{fig:FBC}
\end{figure}

To examine how the proposed methods influence the frequency bias of the network, we measure the spectral bias using the metric--- frequency-band correspondence (FBC) \cite{shi2022measuring}, which first calculates the element-wise $|\mathcal{F}(x)|/|\mathcal{F}(y)|$ between the output $x$ and target image $y$, categorizes it into five frequency bands radially and then computes the per-band averages. Higher values indicate higher correspondence. We trace the evolutions of FBC for $\mathbf{A_8\mathunderscore_{256}}$ and $\mathbf{A_2\mathunderscore_{256}}$ and the corresponding PSNR curves throughout the training iterations in three tasks. 

\begin{figure}[t!]
    \centering
    \includegraphics[width=0.95\linewidth]{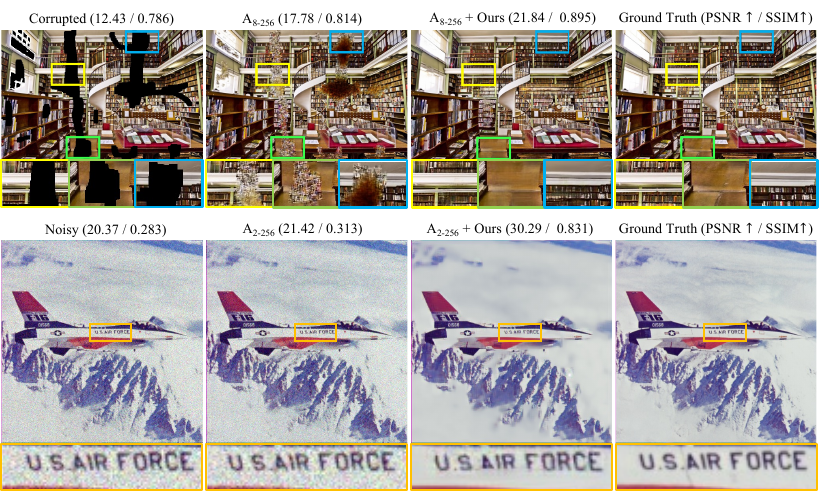}
    \caption{Experiments on natural image inpainting and denoising ($\sigma=25$). Our method improves the extrapolation and denoising capabilities of underperforming architectures.}
    \label{fig:inpainting_denoising}
\end{figure}

\cref{fig:FBC} shows that the original underperforming architectures tend to fit all frequencies more readily, including high frequencies. This is more evident in $\mathbf{A_2\mathunderscore_{256}}$, corresponding to its worst performance among all compared architectures. Our methods substantially delay the convergence of higher frequencies for all three tasks as designed, which decouples the learning of different frequencies. This leads to prolonged denoising effects and enhanced performance for inpainting tasks, including MRI reconstruction, as qualitatively shown in \cref{fig:inpainting_denoising}. We speculate that a stronger bias towards lower frequencies helps the model leverage better the spatial information which improves its inter/extrapolation capability.  Given that MRI reconstruction resembles an inpainting task for $k$-space measurements, we expect similar improvements in natural image inpainting shown in \cref{fig:inpainting_denoising} with our regularized network priors for $k$-space interpolation in MRI experiments. 

\section{Conclusion}
We introduce efficient, architecture-agnostic methods for frequency control over the network priors, offering a novel solution to simultaneously address the key challenges present in untrained image reconstruction. Our approach requires only minimal modifications to the original DIP scheme while achieving significant gains in accruacy and efficiency as evidenced in MRI reconstruction and natural image restoration tasks, making it a stronger zero-shot reconstructior with the potential for seamless integration with other advancements in self-supervised reconstruction.


\section*{Acknowledgements}
This work was supported in part by the United States National Institutes of Health (NIH) through grants R01CA266702 and R01EB035160.

%
%
\bibliographystyle{splncs04}
\bibliography{main}
\end{document}